\documentclass[aps,preprint]{revtex4}%
\usepackage{amsfonts}
\usepackage{amsmath}
\usepackage{amssymb}
\usepackage{graphicx}
\usepackage{slashed}
\usepackage{appendix}%
\begin{document}
\title[Mass shell]{The mass shell of strong-field quantum electrodynamics}
\author{H. R. Reiss}
\affiliation{Max Born Institute, 12489 Berlin, Germany}
\affiliation{American University, Washington, DC 20016-8058, USA}


\begin{abstract}
It has long been known that a free electron in an intense plane-wave field has
a mass shell that differs from the usual free-electron mass shell, with a form
that implies that an intensity-dependent increase in mass occurs. It has been
an enticing, but elusive goal to observe this mass shift. Many schemes have
been proposed by which a definitive measurement may be made, and some claims
of success exist, but these tests are not conclusive. It is shown here that
the intense-field mass shell is not the result of a change in mass. Rather, it
is a consequence of the potential energy that a charged particle must possess
in the presence of a plane-wave field. When the effects of this potential are
incorporated in a properly covariant form, the mass shift no longer appears
and kinematical relations are conventional. If the plane-wave pulse is
sufficiently long to allow the electron to exit the field adiabatically, then
there is no alteration at all of the mass shell expression. Other aspects of
the role played by the ponderomotive 4-potential are examined. It is also
shown that the putative \textquotedblleft relativistic mass\textquotedblright%
\ of the electron is illusory when confronted with covariance requirements.
Both \textquotedblleft mass increases\textquotedblright\ of the electron are
thereby discredited by fundamental principles.

\end{abstract}
\date[17 October 2013]{}
\maketitle

\section{Introduction}

It has been known\cite{sengupta, hr62a,hr62b,nr,goldman,bk} for a half-century
that a charged particle (hereafter referred to generically as an
\textquotedblleft electron\textquotedblright) immersed in an intense
plane-wave field exhibits an intensity-dependent alteration of its mass shell
condition. The field-free mass shell of ordinary quantum electrodynamics (QED)%
\begin{equation}
p^{\mu}p_{\mu}=\left(  mc\right)  ^{2} \label{a}%
\end{equation}
is replaced by the mass shell of Strong-Field QED (SFQED)%
\begin{equation}
p^{\mu}p_{\mu}=\left(  mc\right)  ^{2}\left(  1+z_{f}\right)  , \label{b}%
\end{equation}
where%
\begin{equation}
z_{f}=2U_{p}/mc^{2}. \label{c}%
\end{equation}
The quantity $U_{p}$ is the ponderomotive potential of the electron in the
plane-wave field. The terminology of Ref. \cite{hr62b} is used, since the
conclusion of Ref. \cite{hr62b} that $z_{f}$ is the coupling constant of the
electron to the plane-wave field is relevant here. (In \cite{hr62a,hr62b}, the
$z_{f}$ parameter was designated as $z$ without a subscript.) Equation
(\ref{b}) appears to indicate that the mass of the electron has increased as a
result of interaction with the strong field.

A problem of interpretation arises because Eq. (\ref{b}) can be placed in the
form%
\begin{equation}
p^{\mu}p_{\mu}=\left(  m^{2}+\Delta m^{2}\right)  c^{2}, \label{d}%
\end{equation}
whereas a simple shift in mass $\Delta m$ would lead to%
\begin{equation}
p^{\mu}p_{\mu}=\left(  m+\Delta m\right)  ^{2}c^{2}. \label{e0}%
\end{equation}

A considerable literature has arisen with respect to this revision of the mass
shell condition. An alteration of the mass of the electron that is dependent
on the intensity of the field in which it is immersed has implications for the
foundations of QED. For example, Ref. \cite{kibblesalam} examines the effects
on basic symmetries suggested by an intense-field mass shift.

Within the growing body of literature on the subject of the
intensity-dependent mass, differences of opinion inevitably
arise\cite{mackenroth,corson}. From very early times, ways to observe the mass
shift by laboratory measurements have been suggested\cite{hr66,hr72}. These
suggestions continue to the present day\cite{harvey}. Most of these proposals
amount to a quest for an alteration in kinematics following from the presence
of the ponderomotive potential $U_{p}$. A study of the implications of the
ponderomotive potential is a focus of this article.

Possibly the earliest attempt to measure the mass shift was in
1970\cite{mowat}. Claims have been made\cite{bula,bamber,meyerhofer} that the
mass shift was observed, although these include the caution that the evidence
may not be decisive\cite{bamber}.

A direct approach is taken here to understanding the origin of the
free-electron mass shell condition of SFQED. It is found that it is a
straightforward matter to trace the origins of this alteration, with the
result that one can assign a simple meaning to it. Rather than focusing on
mass \textit{per se}, it is more fruitful to consider energy and momentum
conditions. The result is that the ponderomotive potential provides the
essential key to the explanation of the electron's modified kinematical properties.

The ponderomotive potential has several aspects to its physical significance
that will be explicated after revealing its mass shell implications.

Section II analyses the mass shell condition in terms of the requirement for a
potential energy due to the interaction of a charged particle with the
plane-wave field in which it is immersed. When this potential energy is
explored in covariant terms, it is found to provide a complete explanation for
the modified mass shell condition (\ref{b}) or (\ref{d}).

A close analog to the ponderomotive potential of transverse (i.e. plane-wave)
fields is the ponderomotive energy of an electron in a longitudinal field,
where it is often referred to as a \textquotedblleft quiver
energy\textquotedblright. Despite an apparent equivalence, the two concepts
are fundamentally different, as is explained in Section II.

The quantity $z_{f}$ may be regarded as a dimensionless expression of $U_{p}$.
All investigators of free electrons in strong fields have encountered the
equivalent of the $z_{f}$ parameter, albeit with a multitude of different
notations. Some authors prefer a parameter that is proportional to the
strength of the electric field, apparently seeking an equivalence with
longitudinal field quantities where only the electric field is of
significance. The parameter thus defined is proportional to $z_{f}^{1/2}$. For
reasons explicated in Section III, this reasoning is regarded as misleading.
The quantity $z_{f}$ is the coupling constant of SFQED, and its relation to
the perturbative coupling given by the fine structure constant $\alpha$ is
very instructive. The ponderomotive potential $U_{p}$ and its covariant
extension to a ponderomotive 4-potential lie at the heart of SFQED.

Section IV examines the concept of a so-called \textquotedblleft relativistic
mass\textquotedblright\ that has previously been debunked\cite{okun}. However,
the notion continues in general use, and a simple covariance argument is used
here to show how unphysical that concept actually is. The conclusion is that
neither the intense-field mass shift nor the relativistic mass shift actually exist.

Section V gives a brief summary.

\section{Intense-field mass shell}

\subsection{Ponderomotive 4-momentum}

The structure of the mass shell expressed by Eq. (\ref{b}) is open to an
interpretation quite apart from any change of mass. With the left-hand side
expanded, and Eq. (\ref{c}) inserted on the right-hand side, the expression
becomes%
\begin{equation}
\left(  \frac{E}{c}\right)  ^{2}-\mathbf{p}^{2}=\left(  mc\right)  ^{2}%
+2U_{p}m. \label{f}%
\end{equation}
A simple rearrangement gives%
\begin{equation}
E^{2}=\left(  mc^{2}\right)  ^{2}+2U_{p}mc^{2}+\mathbf{p}^{2}c^{2}. \label{g}%
\end{equation}
The mass shell condition follows from the expression for the minimum value
that $E$ can have. The ponderomotive potential $U_{p}$ is a true potential
energy. If the electron were to emerge adiabatically from the plane wave
field, only then would that energy become a kinetic energy. The electron
cannot exist as a physical particle within the plane wave field unless it
possesses the ponderomotive potential energy $U_{p}$. The electron must
acquire the energy $U_{p}$ from the electromagnetic field. In a
uniform-intensity monochromatic field, photons of that field are
unidirectional, and an amount of field energy $U_{p}$ is associated with a
field momentum $U_{p}/c$ in the direction of propagation$.$ Therefore, the
minimum energy of the electron is, from (\ref{g}) with the minimal momentum
$\left\vert \mathbf{p}\right\vert =U_{p}/c$ inserted,%
\begin{equation}
E_{\min}^{2}=\left(  mc^{2}\right)  ^{2}+2U_{p}mc^{2}+U_{p}^{2}. \label{h}%
\end{equation}
The minimum energy condition is thus%
\begin{equation}
E_{\min}=mc^{2}+U_{p}. \label{i}%
\end{equation}
This is consistent with the hypothesis that the electron in the field must
have, at minimum, the rest energy $mc^{2}$ plus the ponderomotive potential
$U_{p}$. An important additional item of information is that the energy
$U_{p}$ acquired from the field comes with the photon momentum $U_{p}/c$
associated with that amount of field energy.

The analysis\cite{td,hr87} of recent experiments\cite{smeenk} that observed
radiation pressure, confirm the $U_{p}$ energy and $U_{p}/c$ momentum assignments.

An alternative, but equivalent approach also starts with the knowledge that an
electron in a transverse field must possess an energy $U_{p}$. Covariance
requires that an energy must be the time part of a 4-momentum%
\begin{equation}
U^{\mu}:\left(  U_{p},U_{p}\widehat{\mathbf{k}}\right)  ,\label{i1}%
\end{equation}
where $\widehat{\mathbf{k}}$ is a unit 3-vector in the direction of field
propagation. The direction and relative amplitude of the space part of \ Eq.
(\ref{i1}) come from the fact that the \textit{4-potential} $U^{\mu}$ arises
from interaction with the plane-wave field. That is, the 4-potential $U^{\mu}$
is a lightlike 4-vector that is parallel to the propagation 4-vector:%
\begin{equation}
k^{\mu}:\left(  \frac{\omega}{c},\mathbf{k}\right)  ,\label{i2a}%
\end{equation}
with the equivalence%
\begin{equation}
U^{\mu}=\frac{U_{p}}{\omega/c}k^{\mu}.\label{i3}%
\end{equation}
The mass shell is then found from the product%
\begin{equation}
\left(  p^{\mu}+\frac{1}{c}U^{\mu}\right)  \cdot\left(  p_{\mu}+\frac{1}%
{c}U_{\mu}\right)  ,\label{i4}%
\end{equation}
where $p^{\mu}$ is free-particle 4-momentum that satisfies Eq. (\ref{a}).
Carrying out the multiplication indicated in Eq. (\ref{i4}) gives%
\begin{equation}
p^{\mu}p_{\mu}+\frac{2}{c}p^{\mu}U_{\mu}+\frac{1}{c^{2}}U^{\mu}U_{\mu}=\left(
mc\right)  ^{2}+\frac{2U_{p}}{\omega}p\cdot k,\label{i5}%
\end{equation}
using (\ref{i3}) and the fact that $U^{\mu}$ is on the light cone. The
expressions (\ref{i4}) and (\ref{i5}) are Lorentz-invariant, so they must be
true in the frame where $p^{\mu}\rightarrow p_{0}^{\mu}:\left(  mc,\mathbf{0}%
\right)  $, so that%
\begin{equation}
\left(  p^{\mu}+\frac{1}{c}U^{\mu}\right)  \cdot\left(  p_{\mu}+\frac{1}%
{c}U_{\mu}\right)  =\left(  mc\right)  ^{2}+2mU_{p},\label{i6}%
\end{equation}
which corresponds to Eqs. (\ref{b}) and (\ref{c}).

The conclusion is that the free-electron canonical 4-momentum must be
supplemented by the ponderomotive 4-potential because the electron exists in
the presence of a plane-wave field. The intense-field mass shift has now
vanished, since the modified mass shell condition arises entirely from the
known presence of the ponderomotive potential when the electron is in
interaction with the field. The full mass-shell condition (\ref{i6}) contains
no added-mass considerations.

The mass shell condition (\ref{b}) is straightforward. The kinetic energy $T$
is found from%
\begin{align}
T  &  =E-E_{\min}=\sqrt{\left(  mc^{2}\right)  ^{2}+2U_{p}mc^{2}%
+\mathbf{p}^{2}c^{2}}-\left(  mc^{2}+U_{p}\right) \label{j}\\
&  =mc^{2}\sqrt{1+\frac{2U_{p}}{mc^{2}}+\frac{\mathbf{p}^{2}}{m^{2}c^{2}}%
}-\left(  mc^{2}+U_{p}\right) \label{k}\\
&  \approx\frac{\mathbf{p}^{2}}{2m}, \label{l}%
\end{align}
where the approximation in the last step (\ref{l}) corresponds to the
nonrelativistic limit%
\begin{equation}
mc^{2}\gg U_{p},\quad m^{2}c^{2}\gg\mathbf{p}^{2}. \label{m}%
\end{equation}
If that nonrelativistic assumption is not justified, then the full expression
(\ref{j}) or (\ref{k}) must be used.

The conclusion is that there is nothing out of the ordinary about the
kinematics. There need not be any allowance for a shifted mass. The appearance
of $U_{p}$ in Eqs. (\ref{j}) and (\ref{k}) is due simply to the well-known
ponderomotive potential of a charged particle in a plane-wave field. The
non-appearance of $U_{p}$ in Eq. (\ref{l}) means that the ponderomotive
potential of a free-particle interaction is very difficult to observe in a
nonrelativistic situation.

There is, however, the caution that return of the ponderomotive energy to the
emergent particle is possible if the laser pulse is sufficiently long that the
electron can exit adiabatically from the field. If $U_{p}$ is returned from
the field to the electron, then so is the associated momentum of magnitude
$U_{p}/c$ in the direction of laser propagation, and the simple mass shell
condition (\ref{a}) would then be recovered. (This long-pulse behavior might
be the cause of the null result of the experiment reported in Ref.
\cite{mowat}, designed to detect the presence of $U_{p}$.)

The conclusion just reached about the ponderomotive potential being the real
source of the SFQED mass shell has not required any alteration of that
expression. Rather, it is a statement than provision for $U_{p}$ \textit{must
}be a part of kinematical considerations. From this point of view, the
putative mass shift is not a supportable concept.

The qualitative puzzle about why the mass shell of SFQED has the form of Eq.
(\ref{b}) or (\ref{d}) and not the simple mass-shift form of Eq. (\ref{e0})
can now be answered. The connection between the ponderomotive energy $U_{p}$
supplied by the field, and the momentum $U_{p}/c$ acquired in that transfer
are related by the zero-mass property of the photon rather than by the
nonzero-mass energy-momentum relationship of the electron. This precludes the
form (\ref{e0}).

\subsection{Lorentz and gauge invariance}

These conclusions are both Lorentz invariant and gauge invariant because the
ponderomotive energy $U_{p}$ is both Lorentz invariant and gauge
invariant\cite{hrjmo}.

Lorentz invariance follows immediately from the fact that $U_{p}$ arises from
the product $A^{\mu}A_{\mu}$ of the 4-vector potential of the field with
itself, as is evident from the defining relation%
\begin{equation}
U_{p}=\frac{e^{2}}{2mc^{2}}\left\langle \left\vert A^{\mu}A_{\mu}\right\vert
\right\rangle , \label{n}%
\end{equation}
where the angle brackets denote a cycle average and the absolute value
brackets are necessary because $A^{\mu}$ is a spacelike 4-vector. Gauge
invariance of $A^{\mu}A_{\mu}$ is less obvious, but it can be
shown\cite{hrjmo} to follow from the requirement for a plane wave field that
dependence on the spacetime 4-vector $x^{\mu}$ can only be in the form of the
covariant phase $k^{\mu}x_{\mu}$, where $k^{\mu}$ is the propagation 4-vector
of the plane wave field. This requirement is imposed as an ansatz by
Schwinger\cite{schwinger} and by Sarachik and Schappert\cite{ss}, and is shown
to be a necessity in Ref. \cite{hrjmo}.

Gauge invariance is so easily demonstrated and so basic that the proof is
replicated here from Ref. \cite{hrjmo}. Since $A^{\mu}$ can depend on $x^{\mu
}$ only in the form of the covariant phase $k^{\mu}x_{\mu}$, then the
generating function $\Lambda$ of the gauge transformation must also have that
property, giving%
\begin{equation}
A^{\mu}\rightarrow\widetilde{A}^{\mu}=A^{\mu}+\partial^{\mu}\Lambda=A^{\mu
}+\left(  \partial^{\mu}\varphi\right)  \frac{d}{d\varphi}\Lambda=A^{\mu
}+k^{\mu}\Lambda^{\prime}, \label{n1}%
\end{equation}
where%
\begin{equation}
\varphi\equiv k^{\mu}x_{\mu}, \label{n2}%
\end{equation}
and $\Lambda^{\prime}$ is the total derivative of $\Lambda$ with respect to
$\varphi$. The inner product of $\widetilde{A}^{\mu}$ with itself is thus%
\begin{equation}
\widetilde{A}^{\mu}\widetilde{A}_{\mu}=\left(  A^{\mu}+k^{\mu}\Lambda^{\prime
}\right)  \left(  A_{\mu}+k_{\mu}\Lambda^{\prime}\right)  =A^{\mu}A_{\mu},
\label{n3}%
\end{equation}
since $k^{\mu}k_{\mu}=0$, and $k^{\mu}A_{\mu}=0$ for a transverse field.

\subsection{Ponderomotive potential vs. quiver energy}

Some further remarks are important for the necessary identification of $U_{p}$
as a true potential energy. In nonrelativistic laser physics, $U_{p}$ is often
referred to as a \textquotedblleft quiver energy\textquotedblright\ associated
with an oscillatory motion of an electron in an oscillating field. That
identification would mean that $U_{p}$ is a kinetic energy and not a potential
energy. The reason for this is the fact that most nonrelativistic theory and
interpretation is done in terms of the G\"{o}ppert-Mayer gauge (also called
the \textit{length gauge)}, which treats a transverse laser field as if it
were a longitudinal field. In a longitudinal field the potential energy of a
particle of charge $q$ in the field is given by $-q\mathbf{r\cdot E}$ (where
$\mathbf{E}$ is the electric field vector), and not by $U_{p}$. The
ponderomotive potential $U_{p}$ does not exist in the G\"{o}ppert-Mayer gauge
as a true potential. It does occur, but in the guise of a kinetic energy. Its
existence as a kinetic quiver energy arises from the presence of apparent
charge and current sources in the G\"{o}ppert-Mayer gauge that do not actually
exist in the laboratory\cite{hrjpb}. The equations employed\cite{corkum} for
the motion of an electron within the G\"{o}ppert-Mayer gauge predict the
quiver energy, but these equations of motion arise from the virtual sources
that are a necessary adjunct of the G\"{o}ppert-Mayer gauge\cite{hrjpb}. This
is a clear contrast with plane wave behavior. A hallmark of plane waves is
that, once formed, they propagate without input from external sources.

Direct laboratory evidence exists that $U_{p}$ is a potential energy and not a
kinetic energy. In typical short-pulse laser ionization experiments, there is
not sufficient time for the ponderomotive potential to be returned to the
photoelectron before the end of the pulse. (See, for example, the discussion
in Ref. \cite{mohideen}.) Linear polarization spectra would have a minimum at
$U_{p}$ were that energy the kinetic quiver energy, and that fact would be
very noticeable in practical strong-field experiments where $U_{p}$ can be of
the order of the binding energy or greater. That is not what is observed;
linear polarization spectra typically peak near zero energy. This is
consistent with the identity of $U_{p}$ as a potential energy.

\section{$z_{f}$ as coupling constant}

Other properties of the ponderomotive potential are relevant; in particular
the role it plays as the effective coupling constant of charged particles with
very strong electromagnetic fields. The purpose of Ref. \cite{hr62b} was to
explore whether the radius of convergence of SFQED differs from that of QED.
This is important because Dyson demonstrated\cite{dyson} that QED has an
essential singularity at the origin in a complex coupling constant plane,
meaning that all the remarkable successes of QED actually follow from a theory
that is only asymptotic. The findings of Refs. \cite{hr62a} and \cite{hr62b}
are that the fine structure constant $\alpha$, the basic coupling constant of
QED, is replaced in SFQED by the intensity dependent parameter $z_{f}$; and
that the essential singularity at the origin in QED does not appear in SFQED.
However, there are other perturbation-limiting singularities away from the
origin that occur in SFQED in intensity-dependent locations.

A qualitative understanding about the intensity-dependent failure of
perturbation theory comes from the observation that, when $U_{p}$ increases to
the point that the minimum number of photons required to achieve the energy
threshold of the process being studied must index upwards to the next larger
integer, this marks an essential singularity in a complex coupling constant
plane. This phenomenon occurs in both free-particle\cite{hr62b} and bound
particle (see Section IX of \cite{hr80}) processes.

The defining expression for $z_{f}$ in Eq. (\ref{c}) has an alternative
expression as%
\begin{equation}
z_{f}=\alpha\rho\left(  2\lambda\slashed{\lambda}_{C}^{2}\right)  ,\label{e}%
\end{equation}
where $\rho$ is the number of photons per unit volume, $\lambda$ is the
wavelength of the field, and $\slashed{\lambda}_{C}$ is the electron Compton
wavelength. The multiplier $\alpha$ in (\ref{e}) is the fine-structure
constant, the coupling parameter of QED. The effective volume within the
parenthesis in Eq. (\ref{e}) is approximately the volume of a right circular
cylinder of radius $\slashed{\lambda}_{C}$ and length $\lambda$. In other
words, the fine-structure constant $\alpha$ of QED is enhanced in SFQED by the
number of photons contained within the volume of a cylinder with a radius of
an electron Compton wavelength, and extended over a wavelength of the plane
wave field.

This is significant because the coupling constant $z_{f}$ is directly
proportional to $U_{p}$, meaning that $U_{p}$ takes on the additional meaning
of measuring the coupling of the electron to the field, as well as specifying
the potential energy of an electron in a transverse field. The quantity
$z_{f}$ can be regarded as the dimensionless form of $U_{p}$.

A final remark concerns the \textquotedblleft multiple-pole\textquotedblright%
\ structure of Volkov Green's functions in monochromatic beams. The mass shell
condition found there takes the form\cite{hrbaps10,jehr,hrje}%
\begin{equation}
\left(  p^{\mu}-nk^{\mu}\right)  \left(  p_{\mu}-nk_{\mu}\right)  =\left(
mc\right)  ^{2}\left(  1+z_{f}\right)  ,\label{a6}%
\end{equation}
for any integer $n$. This is a significant generalization of Eq. (\ref{b}).
The analysis given above applies only to $n=0$, as well as requiring that
$p^{\mu}$ be replaced by $p^{\mu}+U^{\mu}$. However this is sufficient: only
for $n=0$ is the mass shell condition strictly applicable\cite{hrbaps11}. This
follows from the fact that generalizing a monochromatic field to a wave packet
of plane waves moves all poles except for $n=0$ off the real axis in a complex
representation of the Green's function in momentum space.

\section{Relativistic mass}

The primary purpose of this work has been to explore the \textquotedblleft
intense-field mass shift\textquotedblright. There is a quite different concept
of putative mass change known as \textquotedblleft relativistic
mass\textquotedblright. This can be treated with brevity, so that it is
possible to dismiss in a single paper both long-standing notions of mass
alteration of an electron.

The relativistic mass concept holds that an electron has a rest-frame mass
given by $m_{0}$, and that this is altered to
\begin{align}
m  &  =m_{0}\gamma,\label{o}\\
\gamma &  =1/\left(  1-v^{2}/c^{2}\right)  ^{1/2} \label{p}%
\end{align}
when viewed in a frame moving at velocity $v$ with respect to the rest frame.
This point of view was rejected by Okun\cite{okun}, who protested against a
concept that, among other problems, would require different \textquotedblleft
transverse\textquotedblright\ and \textquotedblleft
longitudinal\textquotedblright\ masses. Nevertheless, the relativistic mass
concept has been stoutly defended\cite{sandin}, largely on the grounds of
convenience in teaching.

A simple alternative view is presented here that is fully in accord with
Okun's conclusions. The argument is based on elementary notions of covariance.

A Lorentz vector is defined as any quantity that transforms under a Lorentz
transformation according to the same rule as the basic Lorentz spacetime
vector $x^{\mu}$. A relativistic velocity that is a simple vector follows from
a derivative with respect to proper time $\tau$:%
\begin{equation}
u^{\mu}=\frac{d}{d\tau}x^{u}. \label{q}%
\end{equation}
The relativistic momentum is then just the product of $u^{\mu}$ with the
Lorentz scalar mass $m$:%
\begin{equation}
p^{\mu}=mu^{\mu}. \label{r}%
\end{equation}
No subscript is required for $m$ since it is a Lorentz scalar that represents
a unique property of the electron. By construction, as shown in Eqs. (\ref{q})
and (\ref{r}), $u^{\mu}$ and $p^{\mu}$ are obviously Lorentz vectors.

Confusion becomes possible when the relativistic velocity $u^{\mu}$ is written
in terms of a nonrelativistic velocity. Since time undergoes an apparent
dilation in any frame other than the rest frame, one has the connection%
\begin{equation}
t=\gamma\tau,\label{s}%
\end{equation}
so that Eq. (\ref{q}) can be expressed as%
\begin{equation}
u^{\mu}=\gamma\frac{d}{dt}x^{\mu},\label{t}%
\end{equation}
which makes it possible to write the momentum 4-vector as%
\begin{equation}
p^{\mu}=m\gamma\frac{d}{dt}x^{\mu}.\label{u}%
\end{equation}
If the factors in Eq. (\ref{u}) are grouped as%
\begin{equation}
p^{\mu}=\left(  m\gamma\right)  \left(  \frac{d}{dt}x^{\mu}\right)  ,\label{v}%
\end{equation}
this makes it possible to introduce the confusing notion of a variable mass
$m\gamma$. When the $\gamma$ factor is employed as in Eq. (\ref{t}) to refer
to a Lorentz vector velocity $u^{\mu}$, the 4-momentum $p^{\mu}$ has the
necessary Lorentz form. The form given in Eq. (\ref{v}) loses covariance
completely by multiplying the noncovariant factor $\left(  m\gamma\right)  $
by another noncovariant factor $\left(  dx^{\mu}/dt\right)  $. This is
needless and confusing.

\section{Summary}

The essential conclusions of this paper can be summarized: The requirement
that an electron in a plane wave field must possess the ponderomotive
potential $U_{p}$ due to that field, coupled with the fact that the
acquisition of $U_{p}$ from the field also gives the electron a minimum
momentum, has been shown to provide a complete explanation for the mass shell
expression of strong-field QED. No change in the mass of the electron occurs.
It is further remarked that the explanation for the existence of the apparent
mass shift in terms of the ponderomotive potential precludes the mass-shift
interpretation. The question of whether Eq. (\ref{b}) might follow from a
shift in mass or from the presence of the ponderomotive potential of an
electron in a plane wave field is resolved in favor of the latter explanation.

The potential $U_{p}$ itself is Lorentz invariant, gauge invariant, and
determines the strength of coupling between a plane-wave field and a charged
particle. Its presence and its effects are fundamental. With that
acknowledgement, the intensity-dependent mass shift hypothesis must be discarded.

The putative variable \textquotedblleft relativistic mass\textquotedblright%
\ of an electron destroys an otherwise straightforward retention of
covariance. It has no redeeming features, and should not be used.

The net conclusion is that both forms of variable electron mass are
unnecessary, and serve only to muddle the underlying physics.

\end{document}